\documentclass[conference]{IEEEtran}
\IEEEoverridecommandlockouts

\usepackage{cite}
\usepackage{amsmath,amssymb,amsfonts}
\usepackage{algorithmic}
\usepackage{graphicx}
\usepackage{textcomp}
\usepackage{xcolor}
\def\BibTeX{{\rm B\kern-.05em{\sc i\kern-.025em b}\kern-.08em
    T\kern-.1667em\lower.7ex\hbox{E}\kern-.125emX}}
\begin{document}

\title{EEG-Based Speech Decoding: A Novel Approach Using Multi-Kernel Ensemble Diffusion Models\\
\thanks{This work was partly supported by Institute of Information \& Communications Technology Planning \& Evaluation (IITP) grant funded by the Korea government (MSIT) (No. RS--2021--II--212068, Artificial Intelligence Innovation Hub, No. RS--2024--00336673, AI Technology for Interactive Communication of Language Impaired Individuals, and No. RS--2019--II190079, Artificial Intelligence Graduate School Program (Korea University)).}
}

\author{
\IEEEauthorblockN{Soowon Kim}
\IEEEauthorblockA{\textit{Dept. of Artificial Intelligence} \\
\textit{Korea University} \\ 
Seoul, Republic of Korea \\
soowon\_kim@korea.ac.kr}

\and

\IEEEauthorblockN{Ha-Na Jo}
\IEEEauthorblockA{\textit{Dept. of Artificial Intelligence} \\
\textit{Korea University} \\
Seoul, Republic of Korea \\ 
hn\_jo@korea.ac.kr}

\and

\IEEEauthorblockN{Eunyeong Ko}
\IEEEauthorblockA{\textit{Dept. of Artificial Intelligence} \\
\textit{Korea University} \\ 
Seoul, Republic of Korea \\
eunyeong\_ko@korea.ac.kr}
}

\maketitle

\begin{abstract}
In this study, we propose an ensemble learning framework for electroencephalogram-based overt speech classification, leveraging denoising diffusion probabilistic models with varying convolutional kernel sizes. The ensemble comprises three models with kernel sizes of 51, 101, and 201, effectively capturing multi-scale temporal features inherent in signals. This approach improves the robustness and accuracy of speech decoding by accommodating the rich temporal complexity of neural signals. The ensemble models work in conjunction with conditional autoencoders that refine the reconstructed signals and maximize the useful information for downstream classification tasks. The results indicate that the proposed ensemble-based approach significantly outperforms individual models and existing state-of-the-art techniques. These findings demonstrate the potential of ensemble methods in advancing brain signal decoding, offering new possibilities for non-verbal communication applications, particularly in brain-computer interface systems aimed at aiding individuals with speech impairments.

\end{abstract}

\begin{IEEEkeywords}
brain--computer interface, electroencephalogram, spoken speech, diffusion model;
\end{IEEEkeywords}

\section{INTRODUCTION}
Speech is a fundamental aspect of human communication, enabling the conveyance of intricate thoughts and ideas \cite{clark1996using}. It is deeply embedded in our social and cultural contexts, playing a critical role in relationship building and information sharing. However, individuals with conditions such as locked-in syndrome are often unable to engage in verbal communication due to physical limitations \cite{laureys2005locked}. Therefore, the development of innovative approaches to restore or replace speech capabilities remains a vital research frontier \cite{han2020classification,brumberg2010brain}. This study focuses on decoding brain signals as a means to facilitate non-verbal communication for such individuals.

Electroencephalography (EEG) provides a non-invasive method to capture the electrical activities of the brain through scalp electrodes \cite{kim2015abstract}. EEG signals have been widely used in applications ranging from neuroscience research to clinical diagnostics \cite{pfurtscheller1999event, prabhakar2020framework}. A growing area of interest involves the decoding of EEG signals to derive meaningful information, such as speech-related activities or cognitive states \cite{teplan2002fundamentals}. EEG-based brain-computer interfaces (BCIs) have been explored for a variety of applications, including mental state classification \cite{lee2020continuous}, emotion recognition \cite{schalk2007decoding}, and motor imagery \cite{mane2020multi}.

Decoding EEG data related to spoken language poses significant challenges due to the complex and highly variable nature of neural activity associated with speech perception and production \cite{kraus2009speech}. EEG signals are also prone to noise and artifacts, which further complicate accurate interpretation \cite{sanei2013eeg,lee2020neural}. As a result, the development of robust and effective methods for EEG decoding is an ongoing area of research with broad applications, including speech rehabilitation and human-machine interfaces \cite{moses2019real}. Previous studies have attempted to decode imagined speech from EEG signals \cite{kim2023diff,lawhern2018eegnet}, demonstrating the potential of EEG-based BCIs for communication.

Deep learning techniques have shown promise in addressing these challenges by automatically learning hierarchical representations from raw EEG data \cite{wang2019deep}. Architectures such as DeepConvNet \cite{schirrmeister2017deep} and EEGNet \cite{lawhern2018eegnet} have been used successfully for EEG decoding tasks \cite{schirrmeister2017deep,lee2023towards}. Other deep learning models, including multi-view CNNs \cite{mane2020multi} and multimodal deep learning networks \cite{han2020classification}, have also been applied to EEG classification tasks, achieving notable success. In addition, graph-based methods have been utilized for EEG analysis to identify patterns in brain networks \cite{yu2019weighted}.

Denoising diffusion probabilistic models (DDPMs) have emerged as powerful tools for learning complex, high-dimensional patterns in data by progressively adding and then removing Gaussian noise \cite{ho2020denoising}. These models have proven effective in dealing with time series data, including audio and video streams \cite{tashiro2021csdi}, making them suitable candidates for EEG signal analysis. Recent studies have applied diffusion-based models to time series data for tasks such as imputation and forecasting \cite{rasul2021autoregressive}. In the context of EEG decoding, diffusion-based models have been explored to decode imagined speech \cite{kim2023diff}.

Building on these approaches, our study aims to further advance the field of EEG-based speech decoding by employing an ensemble learning strategy. We utilize DDPMs combined with conditional autoencoders (CAEs) to capture the intricate neural features associated with spoken speech. By incorporating multiple models with varying kernel sizes, we are able to capture EEG features at multiple temporal scales, thereby improving the robustness and accuracy of the decoding process. Similar multi-scale approaches have been successfully applied in mental state classification \cite{lee2020continuous} and speech-related brain signal analysis \cite{lee2023sentence}.

To our knowledge, this is the first study to apply an ensemble of diffusion models with multi-kernel convolutional layers to decode EEG signals associated with overt speech. By combining the strengths of DDPMs, CAEs, and ensemble learning, we aim to significantly improve the performance of EEG decoding for non-verbal communication, with promising implications for BCI systems that assist individuals with speech impairments.

\begin{figure}[t]
  \centering
  \includegraphics[width=\linewidth,height=\textheight,keepaspectratio]{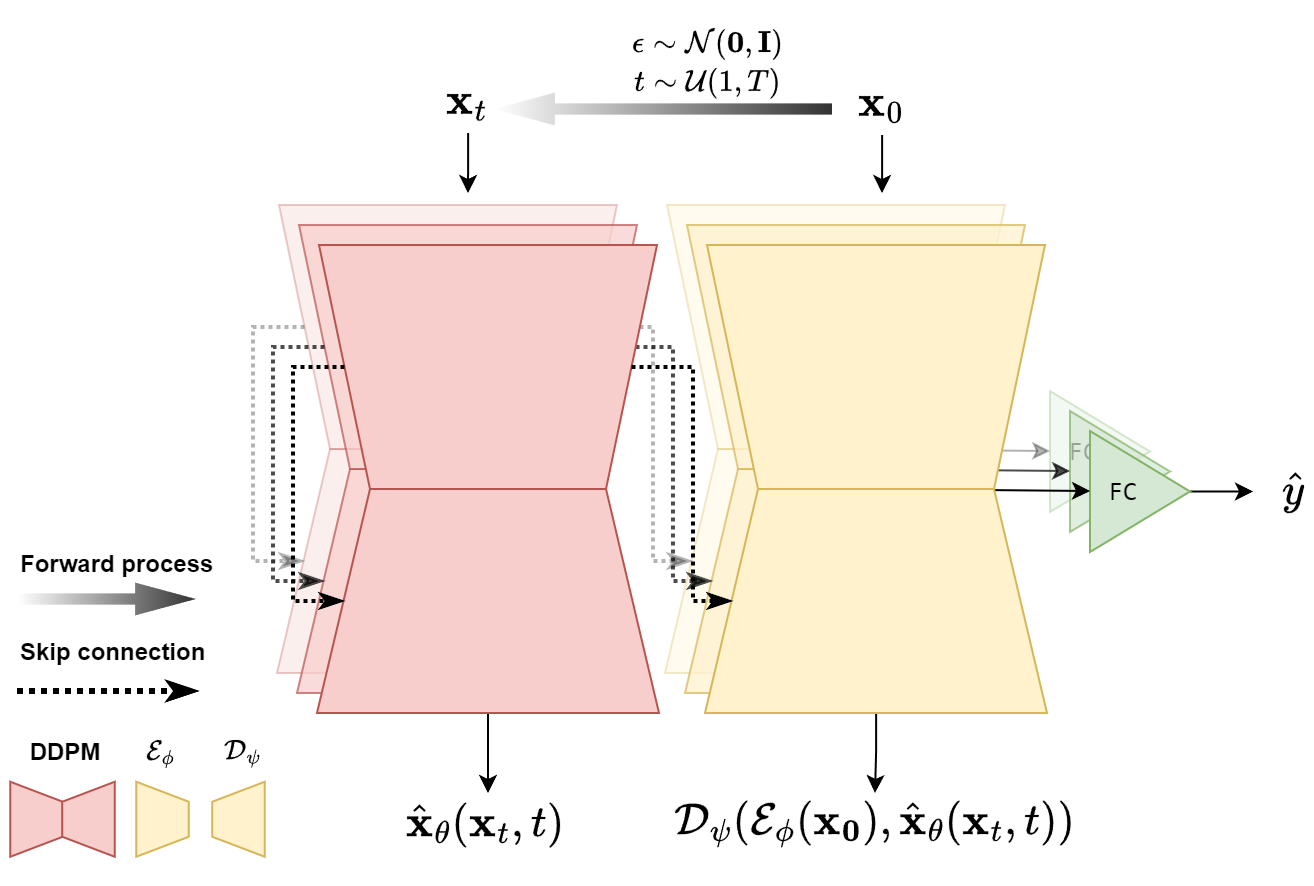}
  \caption{Flowchart of the proposed EEG signal decoding method using an ensemble of models with different kernel sizes (51, 101, and 201). The DDPM first iteratively refines noisy EEG data toward an approximation of the original signal. Each model in the ensemble processes the data with different convolutional kernel sizes to capture multi-scale features. The refined signals are then passed to a CAE for further enhancement. The fully connected (FC) classifier utilizes the ensemble output from the CAE's encoder for downstream tasks such as phoneme or word classification, improving the overall decoding accuracy.}
  \label{fig:method}
\end{figure}

\section{MATERIALS AND METHODS}
The proposed method utilizes an ensemble of DDPMs to effectively capture the multi-scale temporal features of EEG signals. Each model in the ensemble is configured with a different convolutional kernel size—specifically, kernel sizes of 51, 101, and 201—to analyze temporal dependencies at various scales, as depicted in Fig. \ref{fig:method}. This multi-scale approach allows the system to capture both fine-grained and coarse-grained temporal features inherent in EEG data.

\subsection{Denoising Diffusion Probabilistic Models}

The ``forward process" in DDPMs is determined by a fixed Markov chain that progressively adds Gaussian noise to the data. The process starts with the original uncorrupted data, denoted as $q(\mathbf{x}_0)$, and transforms it using a sequence of Markov diffusion kernels, $q(\mathbf{x}_t|\mathbf{x}_{t-1})$, which are Gaussian with a fixed variance schedule $\{\beta_t\}_{t=1}^T$. This process can be expressed as:

\begin{equation}
q(\mathbf{x}_t|\mathbf{x}_{t-1})=\mathcal{N}(\mathbf{x}_t;\sqrt{1-\beta_t}\mathbf{x}_{t-1},\beta_t\mathbf{I}),
\end{equation}
\begin{equation}
q(\mathbf{x}_{1:T}|\mathbf{x}_0)=\prod_{t=1}^{T}q(\mathbf{x}_t|\mathbf{x}_{t-1}).
\end{equation}

Data in any timestep $t$ can also be directly expressed in terms of the original data $\mathbf{x}_0$:

\begin{equation}
q(\mathbf{x}_t|\mathbf{x}_0)=\mathcal{N}(\mathbf{x}_t;\sqrt{\bar{\alpha}_t}\mathbf{x}_0, (1-\bar{\alpha}_t) \mathbf{I}),
\end{equation} 
where $\alpha_t = 1 - \beta_t$ and $\bar{\alpha}_t = \prod_{s=1}^t \alpha_s$.

Each DDPM model in the ensemble aims to denoise the noisy input and generate an output that closely approximates the original signal. We employ a time-conditional U-Net architecture with modifications suitable for EEG data. Each model predicts a version of the original signal, denoted as $\hat{\mathbf{x}}_\theta^{(k)}(\mathbf{x}_t, t)$, where $k \in \{1,2,3\}$ corresponds to the kernel sizes 51, 101, and 201.

\subsection{Conditional Autoencoder}

The forward diffusion process introduces information loss, which is addressed by CAE. The CAE is designed to recognize and correct these errors, resulting in more accurate representations of the original EEG signals. The ensemble setup enhances this process by providing diverse signal reconstructions that the CAE can refine. The objective function for each CAE corresponding to the $k$-th DDPM model is:

\begin{equation}
\mathcal{L}_{\text{CAE}}^{(k)}(\psi^{(k)}, \phi^{(k)}) = \left\| \mathbf{x}_0 - \mathcal{D}_{\psi^{(k)}}\left( \mathcal{E}_{\phi^{(k)}}(\mathbf{x}_t), \hat{\mathbf{x}}_\theta^{(k)}(\mathbf{x}_t, t) \right) \right\|,
\end{equation} 
where $\mathcal{E}_{\phi^{(k)}}$ and $\mathcal{D}_{\psi^{(k)}}$ are the encoder and decoder of the CAE for the $k$-th model.

\subsection{Classifier Ensemble}

After processing through each CAE, the outputs are condensed into latent representations $\mathbf{z}^{(k)}$ using adaptive average pooling layers. Each latent vector is then fed into its corresponding linear classifier $\mathcal{C}_{\rho^{(k)}}$. The predicted labels from each classifier are $\hat{y}^{(k)} = \mathcal{C}_{\rho^{(k)}}(\mathbf{z}^{(k)})$. The final predicted label $\hat{y}$ is obtained by averaging the outputs of the three classifiers:

\begin{equation}
\hat{y} = \frac{1}{3} \left( \hat{y}^{(1)} + \hat{y}^{(2)} + \hat{y}^{(3)} \right).
\end{equation}

The overall objective function combines the reconstruction losses and the classification losses from all three models:

\begin{equation}
\mathcal{L}_{\text{Total}} = \sum_{k=1}^{3} \left( \mathcal{L}_{\text{CAE}}^{(k)}(\psi^{(k)}, \phi^{(k)}) + \alpha \left\| \hat{y}^{(k)} - y \right\|_2 \right),
\end{equation} 
where $\alpha$ is a hyperparameter controlling the balance between reconstruction and classification losses, set to $0.1$ in our experiments.

\subsection{Model Implementation Details}

In our study, we employ an ensemble of three DDPMs with convolutional kernel sizes of 51, 101, and 201, respectively. This design enables the models to capture EEG features at multiple temporal scales, enhancing the ability to model the complex temporal dynamics of EEG signals. Each DDPM and its corresponding CAE consist of convolutional, normalization, and activation layers tailored to effectively process EEG data.

The classifiers $\mathcal{C}_{\rho^{(k)}}$ are trained jointly with their respective CAEs. The latent vector $\mathbf{z}^{(k)}$ for each model has a fixed dimension of 256. Optimization is carried out using the RMSProp optimizer with a cyclic learning rate starting at $9 \times 10^{-5}$ and capped at $1.5 \times 10^{-3}$. Training is carried out over 500 epochs, using L1 loss for the DDPMs and CAEs, and mean squared error for the classifiers’ one-hot encoded outputs.

For model evaluation, 20 \% of the data is reserved for testing, with a consistent random seed to ensure reproducibility. During inference, the predicted labels from the three classifiers are averaged to obtain the final prediction, as described in Equation (7):

\begin{equation}
\hat{y} = \frac{1}{3} \left( \hat{y}^{(1)} + \hat{y}^{(2)} + \hat{y}^{(3)} \right).
\end{equation}

By integrating the outputs of multiple classifiers trained on different temporal scales, the ensemble approach enhances the robustness and accuracy of EEG signal classification. This method effectively leverages the strengths of each model to improve overall performance in decoding EEG signals associated with overt speech.

\subsection{Dataset}
\subsubsection{Data Description}
The data utilized in this study were sourced from a previous investigation conducted by Lee et al. \cite{lee2020neural}. The participants included 22 healthy adults, 15 of whom were male, with a mean age of 24.68 ± 2.15 years. None of the participants had a history of neurological conditions, language disorders, hearing, or vision impairments. Additionally, they refrained from drug use for at least 12 hours prior to the study. All participants had received over 15 years of high-quality English education. For the overt speech task, the 22 subjects were asked to produce 12 different words or phrases, such as ``ambulance," ``clock," ``hello," ``help me," ``light," ``pain," ``stop," ``thank you," ``toilet," ``TV," ``water," and ``yes," along with a resting state condition, creating a total of 13 distinct classes. EEG signals were recorded using a 64-channel cap fitted with active Ag/AgCl electrodes, following the international 10-20 system. The FCz and FPz channels served as the reference and ground electrodes, respectively. EEG data were collected via Brain Vision/Recorder software (BrainProduct GmbH, Germany) and processed using MATLAB 2018a. The impedance of all electrodes was maintained below 10 $k\Omega$. The 22 blocks of 12 words and the resting state were presented to the participants in random order. Each participant contributed 1,300 samples, comprising 100 samples for each category. The study was approved by the Korea University Institutional Review Board [KUIRB-2019-0143-01] and followed the guidelines of the Declaration of Helsinki.

\subsubsection{Preprocessing}
Several preprocessing techniques were applied in this study to enhance the accuracy of the EEG data. First, a bandpass filter was used to retain signals within the 0.5 to 125 Hz range, along with notch filters at 60 and 120 Hz to remove power line interference. Following this, a common average referencing method was employed to further minimize noise. To eliminate artifacts caused by eye movement and muscle activity, automatic methods were utilized for electrooculography and electromyography removal. Once the artifacts were removed, the EEG signals within the high-gamma frequency band were selected for model training and data analysis. The dataset was segmented into 2-second epochs, with a baseline correction applied 500 ms prior to task onset.

\section{RESULTS AND DISCUSSION}

In this study, we evaluated the performance of our proposed ensemble method, which utilizes three DDPMs with kernel sizes of 51, 101, and 201, against three established models: DeepConvNet \cite{schirrmeister2017deep}, EEGNet \cite{lawhern2018eegnet}, and the approach proposed by Lee et al. \cite{lee2020neural} in the context of decoding EEG signals related to spoken speech. The results, summarized in Table \ref{tab:result}, demonstrate that our ensemble method achieved superior performance in both accuracy and area under the curve (AUC). Specifically, our model obtained an average accuracy of 85.47 \%, with a standard deviation of 4.23 \%, and an average AUC of 97.85 \%, with a standard deviation of 1.67 \%. These results significantly surpass the performance of the baseline methods. DeepConvNet, EEGNet, and the method of Lee et al. \cite{lee2020neural}. achieved average accuracies of 32.34 \%, 42.73 \%, and 57.06 \%, and average AUCs of 73.00 \%, 81.00 \%, and 83.01 \%, respectively, demonstrating the enhanced capability of our proposed ensemble model to effectively decode EEG signals related to speech.

The substantial improvement in performance can be attributed to the ensemble of DDPMs with varying kernel sizes, which allows the model to capture multi-scale temporal features more effectively. By averaging the outputs of classifiers trained on different temporal scales, the ensemble method enhances robustness and generalization, leading to higher classification accuracy and AUC.

An ablation study was conducted to assess the contributions of each component in our model, as shown in Table \ref{tab:ablation}. Removing the DDPMs resulted in a decrease in accuracy to 68.12 \% and AUC to 90.45 \%, indicating the importance of the diffusion models in capturing the complex temporal dynamics of EEG signals. Further removing both the DDPMs and the decoder $\mathcal{D}_{\psi}$ led to a significant drop in performance, with accuracy decreasing to 55.89 \% and AUC to 72.34 \%. This highlights the critical role of both the DDPMs and the CAE in our ensemble framework.

\begin{table}[t]
\centering
\renewcommand{\tabcolsep}{3mm}
\caption{Accuracy and AUC scores for spoken speech classification.}
\label{tab:result}
{
\begin{tabular}{lcc}
\hline
\textbf{Model}     & \textbf{Accuracy (\%)} & \textbf{AUC (\%)}\\ 
\hline
\textbf{DeepConvNet \cite{schirrmeister2017deep}} & 32.34 $\pm$ 5.10 & 73.00 $\pm$ 4.00 \\
\textbf{EEGNet \cite{lawhern2018eegnet}} & 42.73 $\pm$ 3.80 & 81.00 $\pm$ 4.19 \\
\textbf{Lee et al. \cite{lee2020neural}} & 57.06 $\pm$ 6.52 & 83.01 $\pm$ 5.10 \\
\textbf{Proposed Method} & \textbf{85.47 $\pm$ 4.23} & \textbf{97.85 $\pm$ 1.67} \\
\hline
\end{tabular}
}
\end{table}

\begin{table}[t]
\centering
\caption{Ablation Study Assessing the Contributions of Each Component.}
\label{tab:ablation}
{
\begin{tabular}{lcc}
\hline
\textbf{Model Variant}     & \textbf{Accuracy (\%)}  & \textbf{AUC (\%)}\\ 
\hline
\textbf{Proposed Method} & \textbf{85.47 $\pm$ 4.23} & \textbf{97.85 $\pm$ 1.67} \\
\textbf{w/o DDPMs} & 68.12 $\pm$ 5.78 & 90.45 $\pm$ 3.22 \\
\textbf{w/o DDPMs \& $\mathcal{D}_{\psi}$} & 55.89 $\pm$ 6.34 & 72.34 $\pm$ 4.89 \\
\hline
\end{tabular}
}
\end{table}

The experimental results demonstrate that our ensemble approach significantly surpasses existing methods in EEG-based speech decoding. Using multiple DDPMs with varying kernel sizes, the model effectively captures a broader range of temporal features, which is essential for decoding the complex EEG signals associated with speech.

The ablation study underscores the importance of each component in our model. The performance drop when the DDPMs are removed highlights their crucial role in denoising and reconstructing EEG signals. The further decline upon removal of the DDPM and the decoder $\mathcal{D}_{\psi}$ emphasizes the necessity of the CAE to correct for the loss of information from the diffusion process.

\section{Conclusion}

The experimental results of our study indicate that our ensemble approach significantly outperforms existing methods in EEG-based speech decoding. By leveraging multiple DDPMs with varying convolutional kernel sizes, our model is able to capture a wider and more comprehensive range of temporal features inherent in EEG signals. This multi-scale analysis is crucial for decoding the complex and variable nature of EEG signals associated with speech, as it allows the model to effectively interpret both fine-grained and long-range temporal dependencies in the neural data. Furthermore, the ensemble enhances robustness and generalization by combining the strengths of individual DDPMs tuned to different temporal scales, resulting in improved performance metrics compared to state-of-the-art methods. Overall, our study contributes to improving EEG-based decoding methods by introducing a novel ensemble framework. We provide a foundation for future advancements in non-verbal communication systems and highlight the importance of multi-scale temporal analysis in neural signal processing.

\bibliographystyle{IEEEtran}
\bibliography{REFERENCE}

\end{document}